\def\year{2019}\relax
\documentclass[letterpaper]{article} 
\usepackage{aaai19}  
\usepackage{times}  
\usepackage{helvet}  
\usepackage{courier}  
\usepackage{url}  
\usepackage{graphicx}  
\usepackage{amsfonts}
\usepackage{algorithm}
\usepackage[noend]{algpseudocode}
\usepackage{amsmath}
\usepackage{multirow}
\usepackage{array}

\usepackage{tikz}
\usepackage{lipsum}
\newcommand\copyrighttextt{%
  \footnotesize This paper was presented and published at the thirty-first Annual Conference on Innovative Applications of Artificial Intelligence (IAAI-19), which was a track in Association for the Advancement of Artificial Intelligence (AAAI) 2019, Honolulu, Hawaii, USA.}
\newcommand\copyrightnotice{%
\begin{tikzpicture}[remember picture,overlay]
\node[anchor=south,yshift=10pt] at (current page.south) {\fbox{\parbox{\dimexpr\textwidth-\fboxsep-\fboxrule\relax}{\copyrighttextt}}};
\end{tikzpicture}%
}

\begin{document}
%
\title{Artificial Counselor System for Stock Investment}
\author{Hadi NekoeiQachkanloo\thanks{The first three authors contributed equally to this work.}, Benyamin Ghojogh\footnotemark[1], Ali Saheb Pasand\footnotemark[1], Mark Crowley\\
Department of Electrical and Computer Engineering, University of Waterloo, Waterloo, ON, Canada\\
\{hnekoeiq, bghojogh, ali.sahebpasand, mcrowley\}@uwaterloo.ca \\
}
\maketitle
\begin{abstract}
This paper proposes a novel trading system which plays the role of an artificial counselor for stock investment. In this paper, the stock future prices (technical features) are predicted using Support Vector Regression. Thereafter, the predicted prices are used to recommend which portions of the budget an investor should invest in different existing stocks to have an optimum expected profit considering their level of risk tolerance. Two different methods are used for suggesting best portions, which are Markowitz portfolio theory and fuzzy investment counselor. The first approach is an optimization-based method which considers merely technical features, while the second approach is based on Fuzzy Logic taking into account both technical and fundamental features of the stock market. The experimental results on New York Stock Exchange (NYSE) show the effectiveness of the proposed system. 
\end{abstract}

\copyrightnotice

\section{Introduction}

\begin{figure}[!t]
\centering
\includegraphics[width=3.34in]{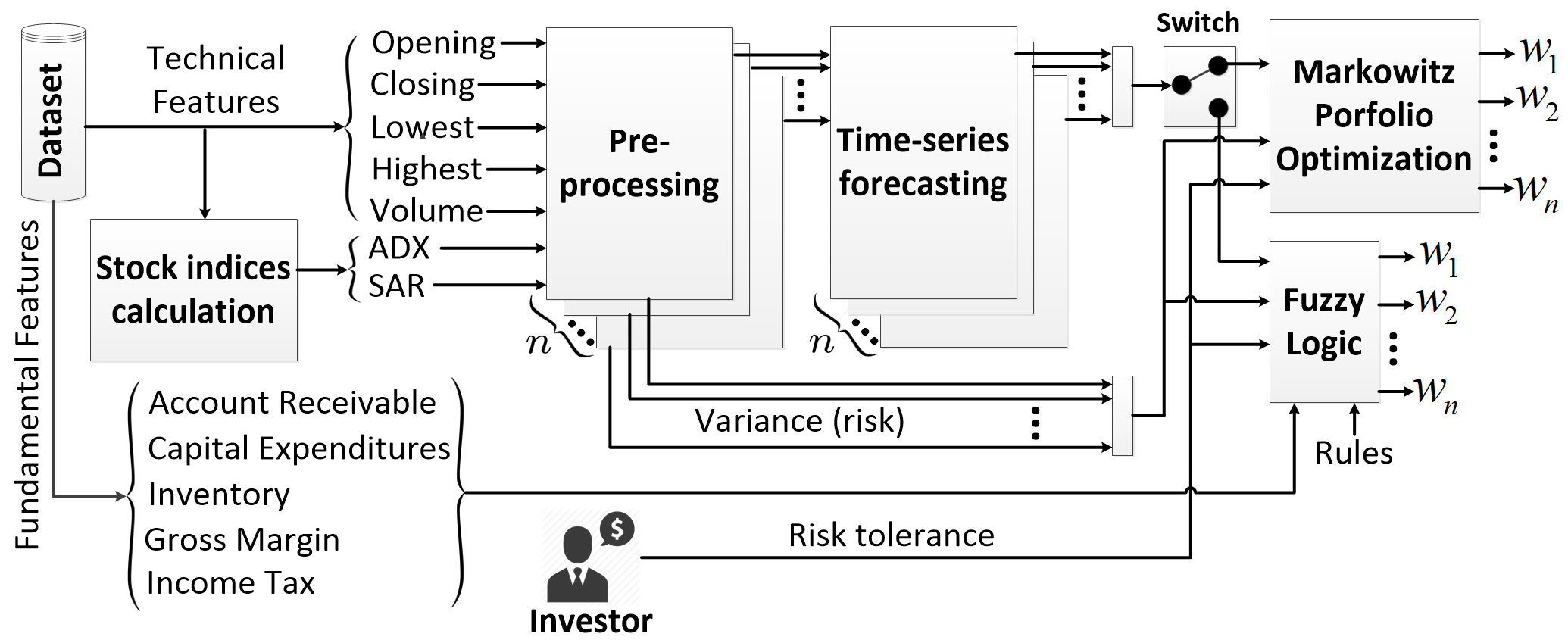}
\caption{Overall structure of the proposed system}
\label{figure_overall_structure}
\vspace{-4mm}
\end{figure}

Consider the situation of an investor, with an existing budget, who wants to know how to divide the budget between several existing stocks. 
The fraction of budget invested in stock $i$ is a weight denoted by $w_i$ in range $[0,1]$. 
The problem is to design an artificial financial counselor system for suggesting the optimum weights for investing in the stocks.  
In order to find the optimum weights, the next day prices of the stocks need to be estimated. 
Therefore, this paper firstly tries to predict the future prices of stocks based on their previous behaviour which can be modeled by time series.  
The primary goal of this project is to find the best weights according to the predicted prices and fluctuations of every stock as well as the risk tolerance of the investor. 
The overall structure of the proposed system is depicted in Fig. \ref{figure_overall_structure}.
In general, there are $n$ stocks each of which has five time series of technical features and several time series of fundamental features. 
Two indices, namely Average Directional Index (ADX) and Parabolic Stop and Reverse Index (SAR), are calculated from the technical features of stocks.
The time series are first preprocessed then a time series predictor forecasts the future prices. The variances (risks) of time series and the future prices go to the next stage computing the optimum weights based on risk tolerance of the investor. That stage can be done in two approaches which are Markowitz portfolio theory \cite{bodie2014investments} (an optimization problem with the purpose of maximizing the profit) and fuzzy investment counselor (fuzzy logic). The latter considers both technical and fundamental features providing opportunity for other sentimental features like an expert broker.

Two approaches are commonly used to analyze the financial behaviour of the markets, (I) technical analysis, and (II) fundamental analysis \cite{cavalcante2016computational,atsalakis2009surveying}. 
The former considers technical attributes which are the history of raw prices of stocks. Some economists believe that all the information needed to predict financial behaviour of the market exist in the prices. Fundamental analysis considers the economic factors, such as liabilities, size of the firm, expenses, assets, revenue, etc, affecting the economic movements of market \cite{vanstone2009empirical,abarbanell1997fundamental}. 

There have been developed different machine learning methods for time series prediction \cite{makridakis2018statistical,cavalcante2016computational}.
Two main methods are commonly used for stock time series prediction \cite{cavalcante2016computational}, which are Muli-Layer Perceptron (MLP) (e.g., \cite{ican2017stock,kayal2010neural}) and Support Vector Regression (SVR) (e.g., \cite{cao2003support,chen2010svm}). In this paper, the new version of the New York Stock Exchange (NYSE) dataset (\url{https://www.kaggle.com/dgawlik/nyse/data}), including both technical and fundamental features, is utilized. In the literature, there are several works attempting to forecast the previous versions of the NYSE dataset, such as \cite{atsalakis2009forecasting,halliday2004equity} and the new version of it \cite{song2018stock}. These works mainly focus on trend prediction of stocks. For instance, a neuro-fuzzy controller is proposed in \cite{atsalakis2009forecasting} for price prediction. The thesis \cite{song2018stock} is one of the very recent works on this dataset. Our work has a comparative performance with it in terms of hit rate. A more important advantage of our system against these works is proposing a complete trading system which recommends the optimum weights of investments as well as predicting the stock exchange, while those works merely attempt to forecast the stocks.
Our main contributions are threefold: 
(I) We propose a complete, novel trading system which not only predicts the future stock prices but also suggests the best fractions of budget for investments. According to \cite{cavalcante2016computational}, there is no such thing as a complete and well-established system proposed in the literature. 
(II) We consider both technical and fundamental attributes of the stocks.
(III) We use two perspectives, i.e., optimization and artificial intelligence, to address the trading problem.

\section{Methodology \& Implementation}

\subsection{Preprocessing}

Three types of preprocessing will be applied on the dataset. The first preprocessing for time series prediction is moving average which smooths the small fluctuations of stock helping to better capture the trend \cite{cavalcante2016computational}. We use moving average on technical features, i.e., opening, closing, lowest, and highest prices, and volume time series, $s'(t) = \frac{1}{\Delta_{ma}} \sum_{k=0}^{\Delta_{ma}-1} s(t-k)$,  
where $s(t)$ and $s'(t)$ denote the data and average data over the period $\Delta_{ma}$ days for day $t$, respectively.
The $\Delta_{ma}$ is $50$ in this work to capture both short-term and long-term trend changes \cite{fletcher2012machine}.

The second preprocessing for time series prediction is Z-score normalization which removes the mean and scales the variance to unit. Every stock has several time series, which are opening, closing, lowest, and highest prices, volume, and the ADX and SAR indices.
In each series, let $\widehat{s}(t)$ denote the normalized data in day $t$. We define $\Delta_p$ as the size of normalization window. 
The normalization is
$\widehat{s}(t) = \big( s'(t) - \overline{s'(t)}\big) /\text{std}\big(s'(t)\big)$,
where $\overline{s'(t)}$, and $\text{std}\big(s'(t)\big)$ are mean and standard deviation of averaged time series in range $[t-\Delta_p, t]$, respectively. 
Note that after time series forecasting, data is denormalized by reversing the transformation. 

Furthermore note that, in economics, the profit rate (relative change of budget) is mostly used rather than the raw budget. Therefore, for the input of weight suggestion modules, the averaged time series is converted to profit rates by
$
r_i(t) = \big(s'(t+1) - s'(t)\big)/s'(t),
$
for every day indexed by $t$ in stock $i$.
Henceforth, $r_i(t)$ is denoted by $r_i$ for simplicity, where $r_i$ means the return of the target future day. Note that the return implies the meaning of profit rate in this work.

\subsection{Prediction of Future Prices}


It has been shown that usage of stock indices is helpful in prediction of individual stock prices \cite{song2018stock}. In this work, two well-known stock indices, Average Directional Index (ADX) and parabolic Stop and Reverse (SAR) \cite{wilder1978new}, are calculated and used for prediction as well as the technical features. In the following, calculation of these two indices are explained and the prediction is detailed.

\subsubsection{Average Directional Index}

The ADX index measures the strength of stock trend regardless of its direction \cite{wilder1978new}. For its calculation, first True Range (TR) is found by
$\text{TR}(t) = \max \big( H(t) - L(t), 
|H(t) - C(t-1)|, |L(t) - C(t-1)|\big),
$
where $H$, $L$, and $C$ denote the highest, lowest, and closing prices of a stock respectively. The smoothed TR (STR) is $\text{STR}(t) = $
\begin{align}\label{equation_smooth_index}
\left\{
    \begin{array}{ll}
        \text{invalid} & \text{if  } t \leq \tau \\
        (1/\tau)\sum_{i=1}^{\tau} \text{TR}(i) & \text{if  } t=\tau+1 \\
        \frac{1}{\tau} \big[ (\tau-1) \text{STR}(t-1) + \text{TR}(t) \big] & \text{otherwise},
     \end{array}
\right.
\end{align}
where $\tau$ is the smoothing period and is set to 14 according to \cite{wilder1978new}.
The Plus Directional Movement (PDM) and Minus Directional Movement (MDM) measure trend direction over time and are calculated as $\text{PDM} = \max \big(H(t) - H(t-1), 0\big)$ and $\text{MDM} = \max \big(L(t-1) - L(t), 0\big)$.
The smoothed PDM (SPDM) and smoothed MDM (SMDM) are calculated using a similar approach to Eq. (\ref{equation_smooth_index}) but with replacing TR with PDM and MDM, respectively. The Smoothed Plus Directional Indicator (SPDI) and Smoothed Minus Directional Indicator (SMDI) are then calculated as $\text{SPDI}(t) = 100 \times \frac{\text{SPDM}(t)}{\text{STR}(t)}$ and $\text{SMDI}(t) = 100 \times \frac{\text{SMDM}(t)}{\text{STR}(t)}$.
Afterwards, the Directional Index (DX) is found as $\text{DX} = 100 \times \frac{\big| \text{SPDI}(t) - \text{SMDI}(t) \big|}{\text{SPDI}(t) + \text{SMDI}(t)}$.
The Average DX (ADX) is obtained using a similar approach to Eq. (\ref{equation_smooth_index}) by replacing TR with DX and $\tau$ with $2\tau$ since $t \leq \tau$ is already invalid for calculating SPDM and SMDM.

\subsubsection{Parabolic Stop and Reverse Index}

The SAR index captures the overall trend of stock over the time. It includes information of stop/reverse points where the trend is changed from uptrend (UT) to downtrend (DT) or vice versa \cite{wilder1978new}. 
The SAR index is $\text{SAR}(t) = $
\begin{align}
\left\{
    \begin{array}{ll}
        \text{invalid} & \text{if  } t < 4 \\
        \min \{L(t-i) | i=0,\dots,3\} & \text{if  } t=4, \text{UT} \\
        \max \{H(t-i) | i=0,\dots,3\} & \text{if  } t=4, \text{DT} \\
        \text{SAR}(t-1) + \text{AF}(t-1) \times \\ \quad  \big( \text{EP}(t-1) - \text{SAR}(t-1) \big) & \text{otherwise},
     \end{array}
\right.
\end{align}
where EP is the Extreme Point and found by $\text{EP}(t) =$
\begin{align}
\left\{
    \begin{array}{ll}
        \text{invalid} & \text{if  } t < 4 \\
        \max \{H(t-i) | i=0,\dots,3\} & \text{if  } t \geq 4, \text{UT} \\
        \min \{L(t-i) | i=0,\dots,3\} & \text{if  } t \geq 4, \text{DT}, \\
     \end{array}
\right.
\end{align}
and AF is the Acceleration Factor and starts from 0.02 and is increased by a step of 0.02 whenever the EP changes. Note that the AF is saturated to 0.2 if it reaches 0.2 and it is reset to 0.02 when the trend changes from UT to DT or vice versa. 

\begin{figure}[!t]
\centering
\includegraphics[width=2.5in]{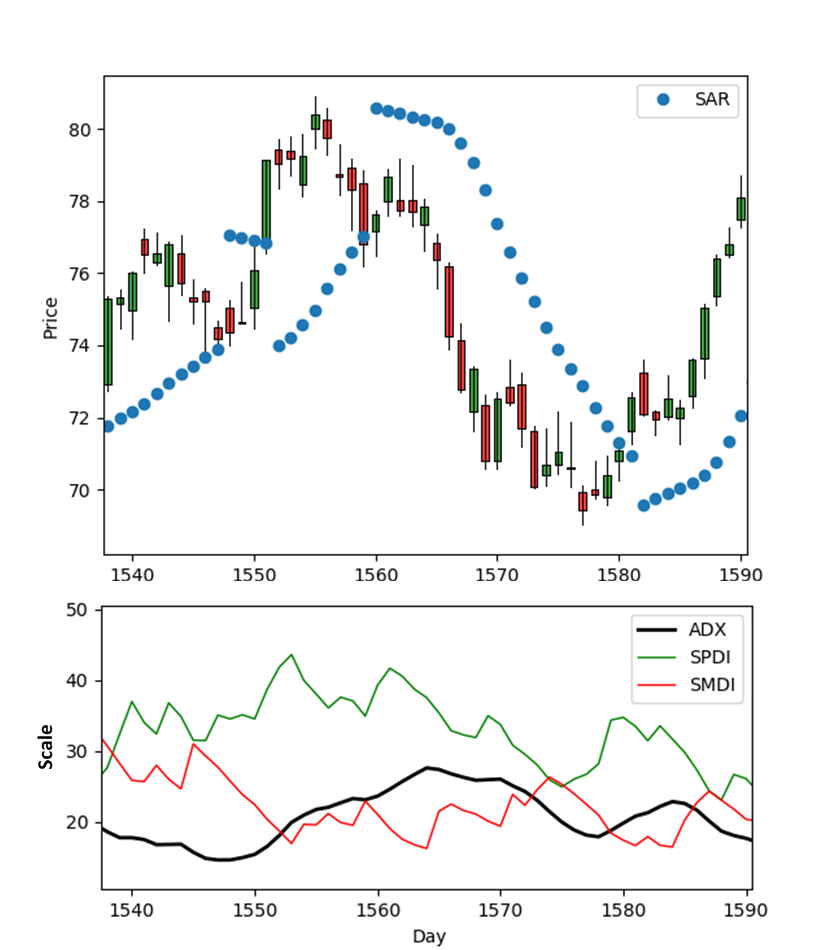}
\caption{The candle plot of a stock and its indices.}
\label{figure_candle_plot}
\vspace{-4mm}
\end{figure}

\subsubsection{Prediction}

To have a visual sense of technical features (opening, closing, highest, and lowest prices) and the SAR, SPDI, SMDI, and ADX indices, see Fig. \ref{figure_candle_plot}. This figure depicts the candle plot of a stock in a time span. The candle plot is a useful visualization of the opening, closing, highest, and lowest prices as well as the trend of stock (green for increasing and red for decreasing). The end points of every candle show the highest and lowest prices of the day and the middle points determine the opening and closing prices. The SAR index is also shown in Fig. \ref{figure_candle_plot}. It can be seen that this index is showing the trend of the stock over time. The SPDI and SMDI indices from which ADX index is found are also depicted. The ADX index is capturing the strength of the stock trend.  For example in days 1550 to 1570, it is showing that stock trend is strong because the difference of SPDI and SMDI is noticeable. 

The future prices of stocks are to be predicted using a time series forecasting method. 
In this work, SVR is used for prediction of stock time series. The utilized kernel function is the radial basis function $k(\mathbf{x}_n, \mathbf{x}_m) = \text{exp} (- \gamma ||\mathbf{x}_n - \mathbf{x}_m||_2^2)$, where $\gamma$, which is found to be $0.001$ by validation, determines the bias of trained model, and $\mathbf{x}_n$ and $\mathbf{x}_m$ are training and testing data, respectively. The $C$ variable, controlling the amount of penalizing slack variables, is found to be $1000$ by validation.
In order to prepare the training and testing data, the technical features (opening, closing, highest, and lowest prices and the volume) as well as the ADX and SAR indices throughout the days within a window with size $\Delta_p$ are concatenated to form a vector. 
We remove the first $2\tau$ days from the training data because of invalidity of indices. 
In this work, the highest price of stocks are predicted and used for suggestion of weights, as explained in the next section because the highest price has a significant impact on investors to invest in a stock or not.

\subsection{Suggestion of Weights}

\begin{figure}[!t]
\centering
\includegraphics[width=2.5in]{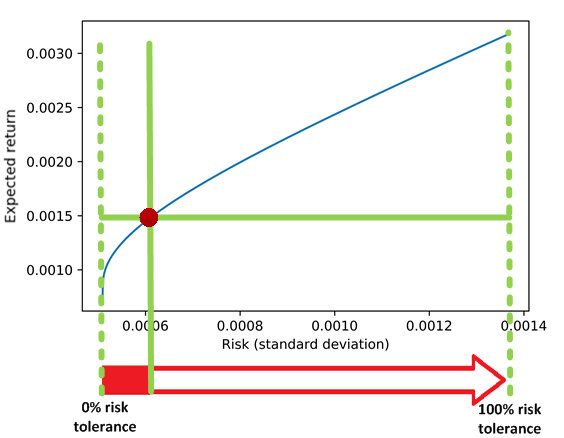}
\caption{The frontier curve of Portfolio theory.}
\label{figure_frontier_curve}
\vspace{-4mm}
\end{figure}

\subsubsection{Markowitz Portfolio Theory}
The first approach for suggesting investment weights is using Markowitz portfolio theory \cite{bodie2014investments}. 
The Markowitz portfolio theory can be formalized as a regularized quadratic optimization problem:
\begin{equation}\label{equation_portfolio_optimization}
\begin{aligned}
& \text{minimize}_{\mathbf{w}}
& & (1 / \mu) \mathbf{w}^\top S \mathbf{w} - \mathbf{w}^\top \mathbf{r} \\
& \text{subject to}
& & \mathbf{w} \succeq 0, \, \, \mathbf{w}^\top \mathbf{1} = 1, \\
\end{aligned}
\end{equation}
where $\mathbf{w} = [w_1, \dots, w_n]^\top$ is the vector of weights whose $i$-th element determines the portion of budget to be invested in stock $i$.
Vector $\mathbf{r} = [\mathbb{E}(r_1), \dots, \mathbb{E}(r_n)]^\top = [\widetilde{r}_1, \dots, \widetilde{r}_n]^\top$ includes the expected value of returns of all the stocks where $\widetilde{r}_i$ denotes the predicted return. The matrix $S$ is the covariance matrix whose element $(i,j)$ is $\mathbb{E}\big[(r_i - \overline{r_i})(r_j - \overline{r_j})\big]$ where $\overline{r_i}$ denotes the mean of $r_i$ in a window of size $\Delta_c$.
The cost function in Eq. (\ref{equation_portfolio_optimization}) consists of two parts, i.e., $\mathbf{w}^\top S \mathbf{w}$ and $\mathbf{w}^\top \mathbf{r}$. The former represents the overall variance (risk) of investment and the latter is the overall expected return of investment. 
The goal of portfolio theory is to maximize the overall expected return of investment while minimizing the overall risk. Note that the weights are positive and they should sum up to one, $\sum_{i=1}^n w_i = 1$.
The $\mu$ in Eq. (\ref{equation_portfolio_optimization}) is the regularization parameter. The higher the $\mu$, the less the variance is penalized and thus the higher risk is allowed to take for investment.

We solved this quadratic optimization problem using Python software for convex optimization (CVXOPT) (\url{https://cvxopt.org/index.html}). The value of $\mu > 0$ is swept from a very small number in order to obtain the frontier curve of portfolio theory \cite{bodie2014investments} which is a curve of expected return versus standard deviation (risk) of investment (see Fig. \ref{figure_frontier_curve}). This sweeping can be stopped whenever the curve reaches the maximum possible risk of investment which is equal to the risk of the stock having maximum variance. In other words, whenever $\mathbf{w}^\top S \mathbf{w} = \max(\mathrm{diag}(S))$.


The maximum risk tolerance of the investor, denoted by $\eta \in [0,1]$, is translated to the range between minimum and maximum standard deviations (risks) of the generated frontier curve of portfolio (see Fig. \ref{figure_frontier_curve}).
The point on the curve having the largest expected return up to that risk level is considered as the optimum portfolio point corresponding to a specific $\mathbf{w}$. 
For calculating $S$, the $\Delta_c$, which is $100$ in this work, should be large enough to capture the true risk of the stock but not so large as to include old and invalid risks.

\begin{figure}[!t]
\centering
\includegraphics[width=2.2in]{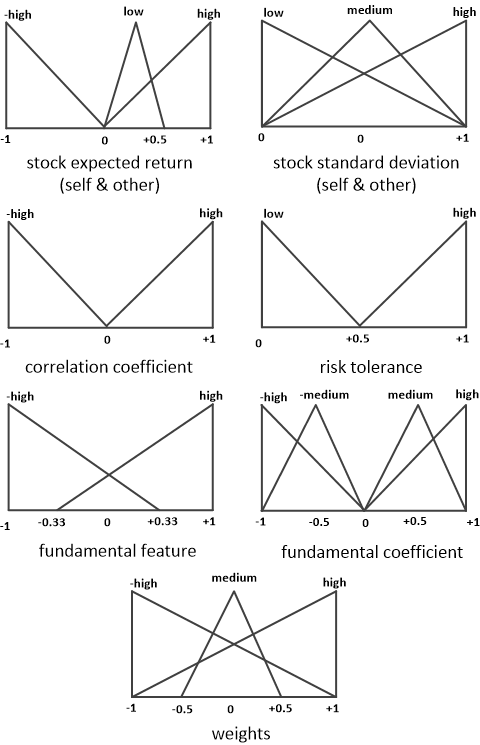}
\caption{Membership functions of fuzzy variables.}
\label{figure_membershipFunctions}
\vspace{-4mm}
\end{figure}

\begin{figure*}[!t]
\centering
\includegraphics[width=7in]{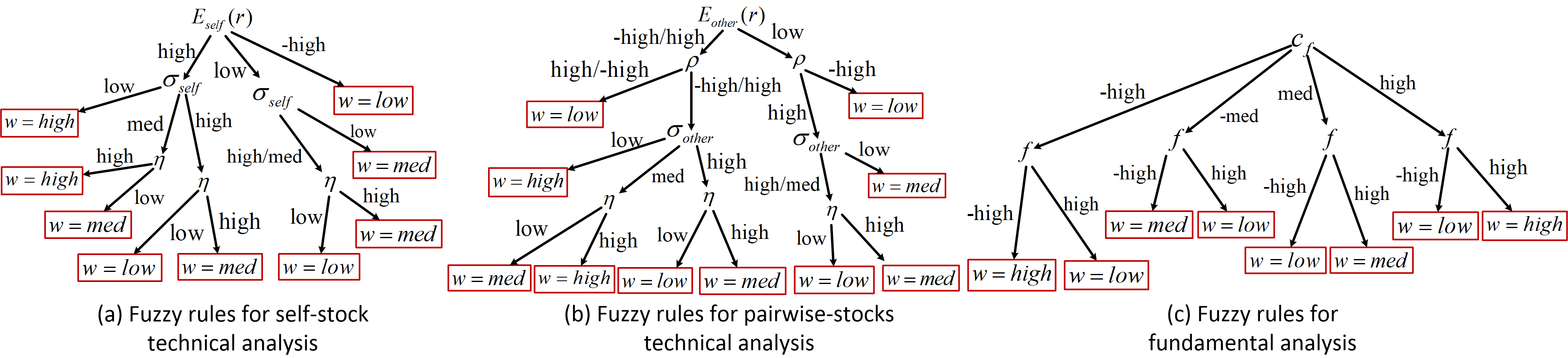}
\caption{Fuzzy rules for (a) self-stock technical analysis, (b) pairwise-stocks technical analysis, and (c) fundamental analysis.}
\label{figure_FuzzyRules}
\end{figure*}


\subsubsection{Fuzzy Investment Counselor}
In this work, a Fuzzy Investment Counselor (FIC) is proposed to model the behaviour of an expert investor under a rationality assumption according to predefined rules. Note that this module takes both technical and fundamental attributes into account while portfolio theory is merely based on technical features. 

\textbf{Parts of FIC:}
The FIC consists of two parts, i.e., technical and fundamental parts.
The technical part includes two separate fuzzy systems. In the technical part, the first fuzzy system, called \textbf{self-stock system}, considers each stock individually, while the second fuzzy system, \textbf{pairwise-stocks system}, considers the effect of all the other stocks on every stock. The fundamental part considers fundamental features of each stock. 

\begin{algorithm}[!t]
\caption{Technical Analysis in FIC}\label{algorithm_fuzzy_technical}
    \begin{algorithmic}[1]
        \State \textit{\textbf{Self-stock Fuzzy system:}}
        \For{$i$ from 1 to $n$} \Comment{iterate on self stock}
            \State $\mathbb{E}'(r_i)$ and $\sigma'_i \gets$ Normalize $\mathbb{E}(r_i)$ and $\sigma_i$
            \State Do fuzzification for $\mathbb{E}'(r_i)$, $\sigma'_i$, and $\eta$ for self-stock fuzzy system
            \State Apply fuzzy rules for self-stock fuzzy system
            \State $w^{t,s}_i \gets$ Do defuzzification for $w$ for self-stock fuzzy system
        \EndFor
        \State \textit{\textbf{Pairwise-stocks Fuzzy system:}}
        \For{$i$ from 1 to $n$} \Comment{iterate on self stock}
            \For{$j$ from 1 to $n$ excluding $i$} \Comment{iterate on other stock}
                \State $\mathbb{E}'(r_j)$ and $\sigma'_j \gets$ Normalize $\mathbb{E}(r_j)$ and $\sigma_j$
                \State Do fuzzification for $\mathbb{E}'(r_j)$, $\sigma'_j$, $\rho_{i,j}$, and $\eta$ for pairwise-stocks fuzzy system
                \State Apply fuzzy rules for pairwise-stocks fuzzy system
                \State $w^{t,p}_{i,j} \gets$ Do defuzzification for $w$ for pairwise-stocks fuzzy system
            \EndFor
            \State $w^{t,p}_i \gets \sum_{j=1}^{n-1} \rho_{i,j} w^{t,p}_{i,j}$
            \State $w^t_i \gets (\eta \times w^{t,p}_{i}) + w^{t,s}_{i}$ \Comment{Fusion of Self-Stock and Pairwise-Stock}
        \EndFor
        \State $\mathbf{w}^t \gets [w^t_1, \dots, w^t_n]^\top$
        \State $\mathbf{w}^t \gets \mathbf{w}^t / \sum_{i=1}^n w^t_i$
  \end{algorithmic}
\end{algorithm}

\begin{algorithm}[!t]
\caption{Fundamental Analysis in FIC}\label{algorithm_fuzzy_fundamental}
    \begin{algorithmic}[1]
        \State $\mathbf{c}_f \gets [0.3, 0.15, -0.4, -0.5, -0.9]^T$
        \For{$i$ from 1 to $n$} \Comment{iterate on stock}
            \For{$k$ from 1 to $n_f$} \Comment{iterate on fundamental features}
                \State $f'_{i,k} \gets$ Normalize $f_{i,k}$
                \State Do fuzzification for $f'_{i,k}$ and $c_f$
                \State Apply fuzzy rules
            \EndFor
            \State $w^{f}_i \gets$ Do defuzzification for $w$
        \EndFor
        \State $\mathbf{w}^f \gets [w^f_1, \dots, w^f_n]^\top$
        \State $\mathbf{w}^f \gets \mathbf{w}^f / \sum_{i=1}^n w^f_i$
  \end{algorithmic}
\end{algorithm}

\textbf{Settings of Fuzzy Logic:}
The Mamdani fuzzy model \cite{karray2004soft} is used for the FIC. 
The membership functions for fuzzy technical and fundamental variables are shown in Fig. \ref{figure_membershipFunctions}. 
The singleton and centroid methods are used for fuzzification and defuzzification, respectively. 
Note that fuzzification converts crisp variables to fuzzy qualitative values, and defuzzification does the reverse \cite{klir1995fuzzy}. 
The fuzzy rules of both parts are illustrated using tree structures in Fig. \ref{figure_FuzzyRules}. 
Every rule in the illustrated tree is a path from root to a leaf where fuzzy T-norm is applied on the non-leaf nodes and the leaf is the output of the rule.
To better understand these trees, two sample rules from Fig. \ref{figure_FuzzyRules}b are explained here: \textit{``IF $E_\text{other}(r)$ is HIGH and $\rho$ is HIGH and $\sigma_\text{other}$ is MEDIUM and $\eta$ is LOW / HIGH, THEN $w$ is MEDIUM / HIGH.''}
The intuition of this rule is that if the correlation of two stocks is high and the return of one of them is going up with medium risk, it is better to invest in the other stock. However, it depends on the risk tolerance of the investor to give a large or medium weight to investing in that stock. The standard operators (minimum and maximum respectively for T-norm and S-norm) are used in this work.

\textbf{Self-Stock Fuzzy System:}
As seen in Algorithm \ref{algorithm_fuzzy_technical}, the self-stock fuzzy system in the technical analysis part of FIC iterates on the stocks and for each of them, it suggests the weights ($w^{t,s}_i$ for stock $i$) according to its expected return $\mathbb{E}(r_i)$, risk $\sigma_i$ (standard deviation), and risk tolerance $\eta$ (see Fig. \ref{figure_FuzzyRules}a). 
In order to prepare the inputs for the suitable ranges of defined membership functions, the self-stock expected return and self-stock standard deviation are respectively normalized to $\mathbb{E}'(r_i) \gets \mathbb{E}(r_i) / \big(0.001 + \sum_{i=1}^n |\mathbb{E}(r_i)|\big)$ and $\sigma'_i \gets \sigma_i / \big((\sum_{i=1}^n \sigma_i)(0.0001 + \sigma_\text{scaling}) \big)$ where $\sigma_\text{scaling} = \eta + \big((1 - \eta) \times \max_{i=\{1,\dots,N\}} \sigma_i \big)$. 
This $\sigma_\text{scaling}$ is a function of $\eta$ to emphasize the risk tolerance of the investor.

\textbf{Pairwise-Stocks Fuzzy System:}
In the pairwise fuzzy approach (see Algorithm \ref{algorithm_fuzzy_technical}), every stock is found to be better or worse than the other stocks to invest in. 
For every stock $i$, we iterate on all other stocks, indexed by $j$, and according to the rules of pairwise-stocks system (see Fig. \ref{figure_FuzzyRules}b), the weights $w^{t,p}_{i,j}$ are suggested.
The rules are applied on the fuzzified values of other stocks' expected return $\mathbb{E}(r_j)$, risk $\sigma_j$, risk tolerance $\eta$, and mutual correlation coefficient, $\rho_{i,j} = S(i,j) / \sigma_i \sigma_j$, where $S(i,j)$ is the element $(i,j)$ of covariance matrix. 
The expected return and risk of other stocks should also be normalized as explained before. 
Finally, for every stock, $n-1$ pairwise weights are obtained and they should be fused. 
For this fusion for stock $i$, the $w^{t,p}_{i,j}$ weights are summed up while weighted by their $\rho_{i,j}$ which determines their impact $w^{t,p}_i = \sum_{j=1}^{n-1} \rho_{i,j} w^{t,p}_{i,j}$.

\textbf{Fusion of Self-Stock and Pairwise-Stocks Fuzzy Systems:}
Finally, the total technical weight is $w_i^t = (\eta \times w^{t,p}_{i}) + w^{t,s}_{i}$ where the influence of $w^{t,p}_{i}$ is ignored when $\eta$ is small because when the risk tolerance is low, self-stock variations should be more valued. The vector of total technical weights, denoted by $\mathbf{w}^t = [w^t_1, \dots, w^t_n]^\top$, is then normalized so that the weights sum up to one, $\mathbf{w}^t \gets \mathbf{w}^t / \sum_{i=1}^n w_i^t$.

\textbf{Fundamental Analysis in FIC:}
For fundamental analysis in FIC, $n_f$ (here five) fundamental features, i.e., \textbf{account receivable}, \textbf{capital expenditure}, \textbf{inventory}, \textbf{gross margin}, and \textbf{income tax}, are selected to be used in this work from the dataset. We define fundamental coefficients, denoted by $\mathbf{c}_f$, for the utilized fundamental features. This coefficient is in the range $[-1,1]$ and determines the positive or negative impact of a fundamental feature on the stock's expected return. Considering the analysis reported in \cite{abarbanell1997fundamental}, which reports the relative impacts of the mentioned fundamental features on the expected return of the stock, we define $\mathbf{c}_f = [0.3, 0.15, -0.4, -0.5, -0.9]^\top$ as a vector of fundamental coefficients of the used fundamental features with the same mentioned order. As seen in Algorithm \ref{algorithm_fuzzy_fundamental}, for every stock, separate fuzzy rules are used for different fundamental features (see Fig. \ref{figure_FuzzyRules}c). For every stock $i$, the fundamental features, indexed by $k$ and denoted by $f_{i,k}$, are normalized with summation of their absolute values to be appropriate for the defined membership function, $f'_{i,k} \gets f_{i,k}/(0.001 + \sum_{k=1}^{n_f} |f_{i,k}|)$.
The obtained vector of fundamental weights $\mathbf{w}^f = [w^f_1, \dots, w^f_n]^\top$ is finally normalized to sum up to one, $\mathbf{w}^f \gets \mathbf{w}^f / \sum_{i=1}^n w_i^f$. 

\begin{figure}[!t]
\centering
\includegraphics[width=3.45in]{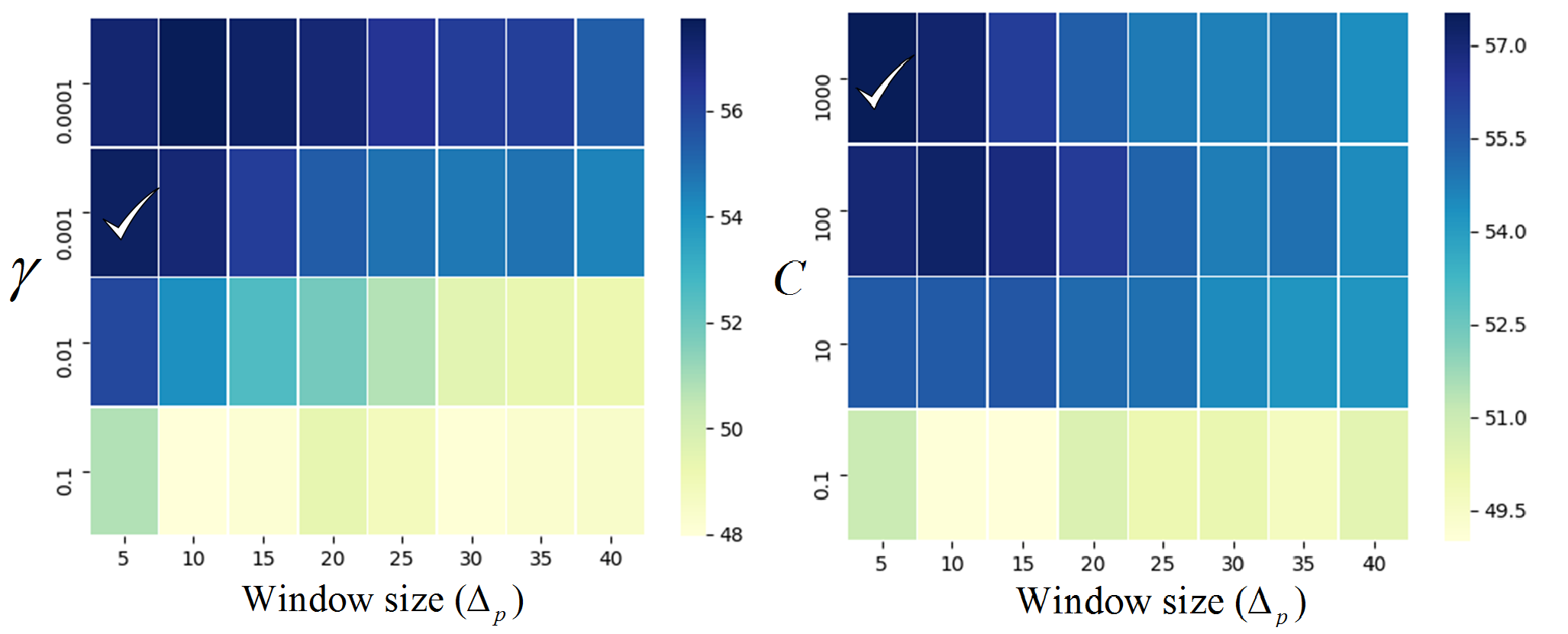} 
\caption{Sweeping $\gamma$ and $C$ versus $\Delta_p$}
\label{figure_parameter_selection}
\vspace{-4mm}
\end{figure}

\textbf{Combining Technical and Fundamental Analysis:}
Eventually, for stock $i$ from technical weight $w_i^t$ and fundamental weight $w_i^f$, the overall fuzzy weight is calculated as $w_i = \alpha w_i^f + w_i^t$ where $\alpha = \big(n_f + \mathbf{c}_f^\top \mathbf{f}'_i\big) / 2 n_f$ and $\mathbf{f}'_i$ is a vector containing the normalized fundamental features of $i$-th stock. Note that $\mathbf{c}_f^\top \mathbf{f}'_i$ is in range $[-n_f, n_f]$ and thus the range of $\alpha$ is $[0,1]$. This $\alpha$ determines the importance of fundamental analysis against the technical analysis. Finally, the overall fuzzy weights $\mathbf{w} = [w_1, \dots, w_n]^\top$ are normalized so that they sum up to one, $\mathbf{w} \gets \mathbf{w} / \sum_{i=1}^n w_i$.

\section{Experimental Results}\label{section_experiments}

\subsubsection{Utilized Dataset}\label{section_datasets}
The utilized dataset in this work is the New York Stock Exchange (NYSE) dataset. 
This dataset contains information on 140 stocks from years 2010 to 2016. We selected 25 well-known stocks for evaluating the prediction and comparison to the literature. Most of the stocks are selected inspired by \cite{song2018stock} although our algorithm can be applied on any number of stocks. 
The 25 selected stocks are AAPL (Apple), AIG (American International Group), AMZN (Amazon), BA (Boeing), CAT (Caterpillar), COF (Capital One), EBAY, F (Ford), FDX (FedEx), GE (General Electric), GM (General Motors), GOOG (Google Alphabet Inc.), HD (The Home Depot), IBM, JNJ (Johnson \& Johnson), JPM (JPMorgan Chase \& Co.), KO (Coca-Cola), MSFT (Microsoft), NKE (Nike), ORCL (Oracle), PEP (PepsiCo), T (AT\&T), WMT (Walmart), XOM (ExxonMobil), and XRX (Xerox).
The used information of this dataset in this work includes:
(I) \textit{Prices:} The prices (in US dollar) represented as time series, which are opening (the morning price), closing (the evening price), lowest, and highest prices as well as volume (the number of traded stocks in the day). For training data, the first 80\% of the data (almost years 2010 to 2014) is used for training and the rest (almost 2015 and 2016) for test. For tuning the parameters, the last 20\% of the training set is held out for validation. 
(II) \textit{Fundamental Attributes:} It contains some numeric metrics showing the annual company performance such as account receivable, capital expenditure, inventory, gross margin, and income tax. Features for years 2013 to 2015 are used as fundamental inputs to FIC. 

\begin{table*}[!t]
\begin{minipage}{\textwidth}
\setlength\extrarowheight{5pt}
\centering
\scalebox{0.6}{    
\begin{tabular}{l | l || c | c | c | c | c | c | c | c | c | c | c | c| c}
\multicolumn{1}{c|}{} &  & \textbf{AAPL} & \textbf{AIG} & \textbf{AMZN} & \textbf{BA} & \textbf{CAT} &  \textbf{COF} & \textbf{EBAY} & \textbf{F} & \textbf{FDX} & \textbf{GE} & \textbf{GM} & \textbf{GOOG} & \textbf{HD} \\
\hline
\hline
\multirow{4}{*}{NSP}  
& HR & \textbf{61.16\%} & 55.36\% & \textbf{57.68\%} & \textbf{60.00\%} & 60.29\% & \textbf{58.26\%} & 56.81\% & 61.45\% & 58.26\% & 53.33\% & 61.46\% & \textbf{57.97\%} & 56.52\%\\
& MAE & 1.058 & 0.410 & 7.973 & 1.310 & 0.848 & 0.757 & 0.291 & 0.131 & 1.517 & 0.228 & 0.332 & 6.676 & 1.086 \\
& RMSE & 1.483 & 0.710 & 11.085 & 1.748 & 1.150 & 1.018 & 0.474 & 0.190 & 2.227 & 0.320 & 0.455 & 9.707 & 1.478 \\
& MAPE & 0.99\% & 0.84\% & 1.23\% & 0.97\% & 1.12\% & 1.04\% & 1.07\% & 1.00\% & 0.97\% & 0.78\% & 1.04\% & 0.92\% & 0.85\% \\
\hline
\multirow{1}{*}{SP}  
& HR & 91.59\% & 92.17\% & 93.33\% & 85.50\% & 88.69\% & 95.65\% & 89.56\% & 86.08\% & 93.04\% & 92.17\% & 83.76\% & 89.70\% & 92.75\% \\
\hline
\multirow{1}{*}{TP \cite{song2018stock}}  
& HR & 57.58\% & 58.79\% & 56.21\% & 60.15\% & $\times$ & 57.12\% & 59.70\% & $\times$ & 59.55\% & 61.21\% & $\times$ & 58.33\% & 59.70\%\\
\hline
\hline
\multicolumn{1}{c|}{} &  & \textbf{IBM} & \textbf{JNJ} &   \textbf{JPM} & \textbf{KO} & \textbf{MSFT} & \textbf{NKE} & \textbf{ORCL} & \textbf{PEP} & \textbf{T} & \textbf{WMT} & \textbf{XOM}  & \textbf{XRX} \\
\hline
\hline
\multirow{4}{*}{NSP}  
& HR & 55.94\% & 52.75\% & 56.23\% & 54.49\% & 53.62\% &  \textbf{60.00\%} & \textbf{58.26\%} & 57.68\% & \textbf{57.68\%} & 56.52\% & 54.49\% & 60.00\%\\
& MAE & 1.178 & 0.613 & 0.598 & 0.236 & 0.461 & 0.598 & 0.286 & 0.579 & 0.229 & 0.502 & 0.669 & 0.110 \\
& RMSE & 1.681 & 0.889 & 0.879 & 0.333 & 0.708 & 0.834 & 0.408 & 0.802 & 0.313 & 0.799 & 0.885 & 0.160 \\
& MAPE & 0.80\% & 0.56\% & 0.92\% & 0.55\% & 0.87\% & 1.02\% & 0.74\% & 0.57\% & 0.61\% & 0.75\% & 0.80\% & 1.10\% \\
\hline
\multirow{1}{*}{SP}  
& HR & 92.46\% & 93.04\% & 94.49\% & 90.14\% & 89.27\% & 90.14\% & 93.91\% & 88.98\% & 93.91\% & 94.49\% & 93.62\% & 85.50\% \\
\hline
\multirow{1}{*}{TP \cite{song2018stock}}  
& HR & $\times$ & 56.52\% & 61.67\% & 59.24\% & 59.09\% & 60.61\% & 58.94\% & 59.39\% & 58.64\% & 60.91\% & 60.45\% & $\times$ \\
\hline
\hline
\end{tabular}%
}
\caption{Evaluation of our work on forecasting technical features and comparing to related work. The $\times$ symbol means that the related work does not have result on that stock.}
\label{table_results_prediction}
\end{minipage}
\end{table*}

\subsubsection{Results for Price Prediction}
The proposed framework includes three parameters $\Delta_p$, $C$, and $\gamma$. To tune these parameters, different values, i.e., $\Delta_p = \{5, 10, \dots, 40\}$, $C = \{0.1, 10, 100, 1000\}$, and $\gamma = \{0.1, 0.01, 0.001, 0.0001\}$ were tested on the validation set. Figure \ref{figure_parameter_selection} illustrates the average hit rates over the 25 stocks on the validation set. 
Note that the Hit Rate (HR) is formulated as $\text{HR} = \frac{1}{m} \sum_{t=1}^m \big[\text{sign}(\widetilde{r}(t)) \stackrel{?}{=} \text{sign}(r(t))\big]$ where $m$ is the number of days of prediction.
The HR assesses how well the trend of price is predicted. As can be seen in Fig. \ref{figure_parameter_selection}, the best obtained parameters are $C=1000$, $\Delta_p=5$, and $\gamma=0.001$. As expected, a large $C$ is penalizing slack variables for better prediction, and small $\gamma$ avoids a biased model. A small window size also makes sense according to \cite{song2018stock}.

\begin{figure}[!t]
\centering
\includegraphics[width=2.6in]{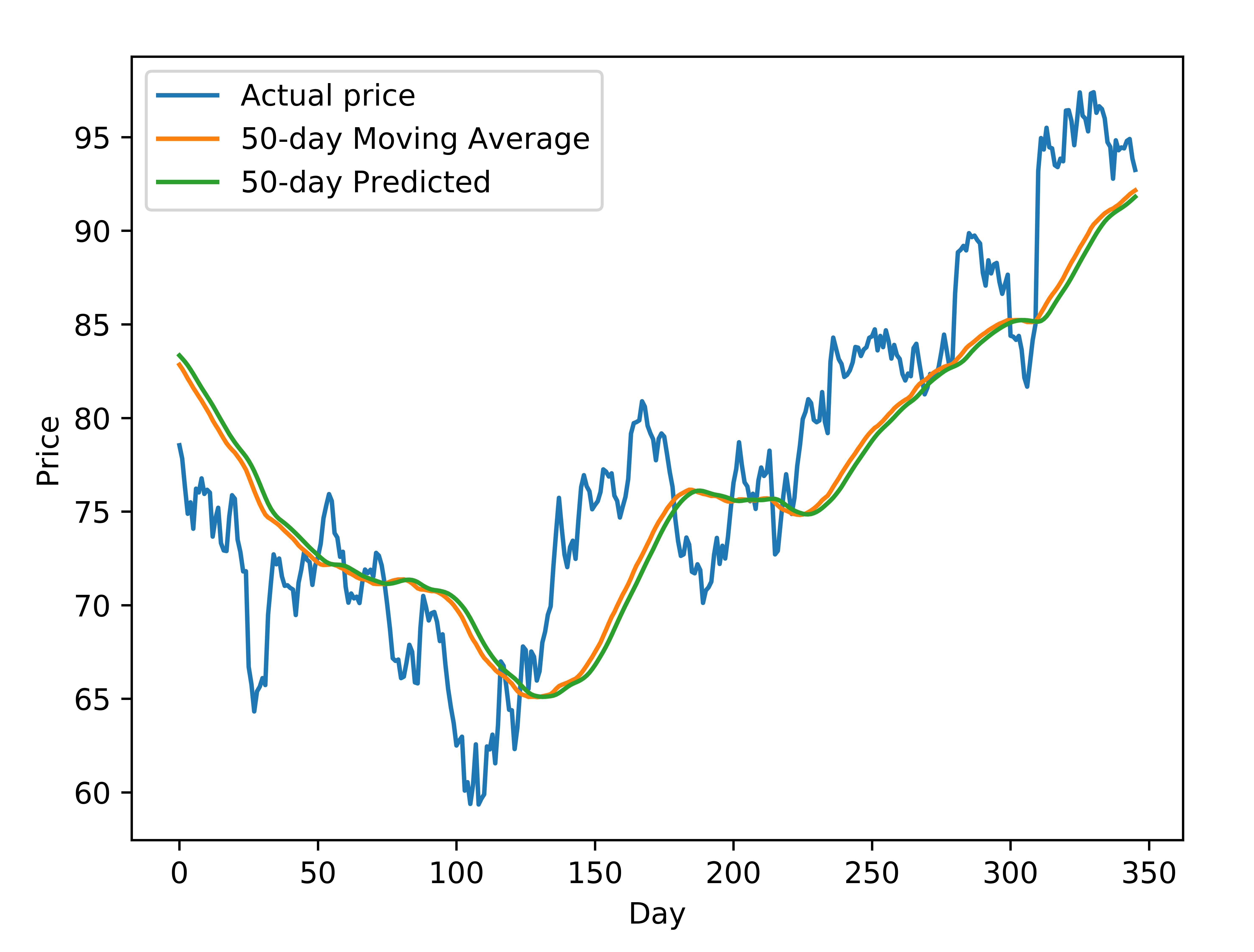}
\caption{Forecasting the averaged time series of highest price in CAT stock.}
\label{figure_prediction}
\end{figure}

We evaluated the predictions using standard methods from the literature, which are Hit Rate (HR), Mean Absolute Error ($
\text{MAE} = (1/m) \sum_{t=1}^m \big|s(t) - \widetilde{s}(t)\big|$), and Root Mean Square Error ($\text{RMSE} = \sqrt{\frac{1}{m} \sum_{t=1}^m \big(s(t) - \widetilde{s}(t)\big)^2}$) where $\widetilde{s}(t)$ denotes the predicted time series. To have a better comparative sense of these metrics, we also report Mean Absolute Percentage Error ($\text{MAPE} = (100/m) \sum_{t=1}^m |\big(s(t) - \widetilde{s}(t)\big) / s(t)|$).
The results on the test set are reported for the 25 stocks in Table \ref{table_results_prediction}. We evaluated price prediction for two cases, i.e., Not Smoothed Prices (NSP) and Smoothed Prices (SP) with moving average, reported in this table.
The MAE, RMSE, and MAPE errors of our work on different stocks are relatively small. Moreover, the hit rates are fairly high in almost all the stocks showing the fine performance of SVR in predicting the time series and its trend. The hit rates of SP are all high enough showing the effectiveness of our algorithm. 
Table \ref{table_results_prediction} also compares the stock prediction of the proposed framework with a very recent work for Trend Prediction (TP) \cite{song2018stock}. As reported in this table, the NSP outperforms TP or has comparative result with it in stocks AAPL, AMZN, BA, COF, GOOG, NKE, ORCL, and T.
It is noteworthy that TP \cite{song2018stock} merely predicts $+1$ for uptrend and $-1$ for downtrend and measures the hit rate based on that. However, NSP predicts the actual price and our hit rate is based on the predicted price (and not the trend) which is more difficult.

In order to capture a visual sense of the performance of price forecasting, see Fig. \ref{figure_prediction} which shows that the predicted averaged time series is following the actual averaged series closely.

\begin{figure}[!t]
\centering
\includegraphics[width=3in]{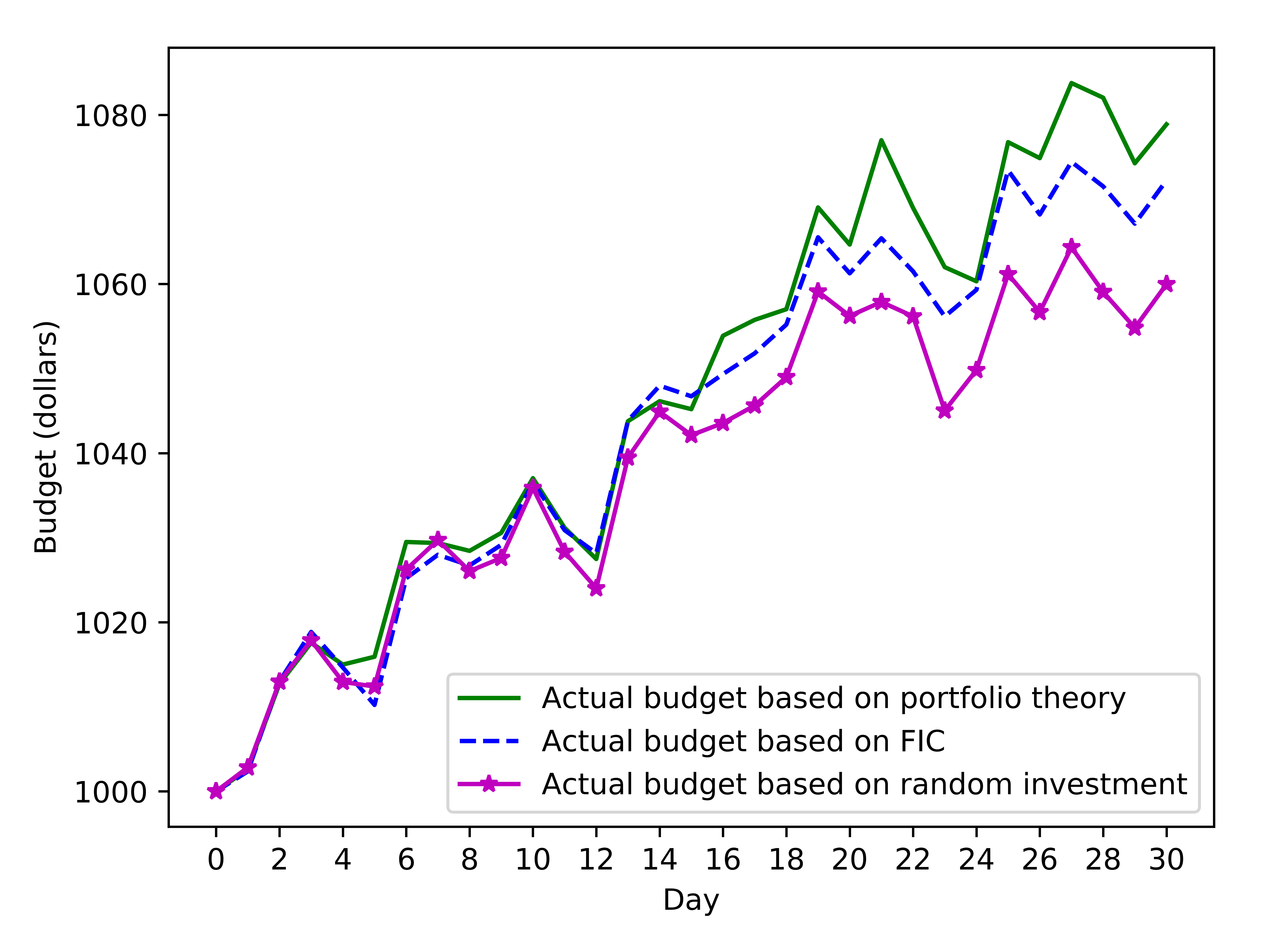}
\caption{Actual and predicted budgets for a 30-day period.}
\label{figure_Budget_plot}
\end{figure}

\subsubsection{Results for Suggestion of Weights}
The predicted averaged prices, i.e., SP, are used as input to weight suggestion part.
In order to illustrate the performance of the weight suggestion modules, suppose that the initial budget of the investor, denoted by $B$, is $\$1000$. 
The proposed weight suggestion methods, which are based on portfolio theory or fuzzy logic (FIC), are tested on this budget invested in the 20 stocks (all except the stocks GE, GOOG, ORCL, and JNJ because of lack of fundamental features in the dataset and the stock GM because of its incomplete time series of technical features in the dataset). 

For every day of a period of time, the optimum weights $\{w_i\}_{i=1}^n$ are found, from which the actual future return $r$ and the expected future return $\mathbb{E}(r)$ are obtained as $\sum_{i=1}^n w_i r_i$ and $\sum_{i=1}^n w_i \mathbb{E}(r_i)$, respectively.
Then the actual future budget ($B$) is updated based on the obtained $r$ as $B \gets B \times (1 + r)$.
Note that, according to rationality, if the expected return is negative, we exit from stock market for that future day.
This procedure is performed for 30 days, as an example, to compare and analyze the actual remained budgets for three different methods of weight suggestion which are portfolio theory, FIC, and random weights. 
This example 30-day period includes days 1537 to 1567 of dataset although can be any time span. 
The actual budgets, are illustrated in Fig. \ref{figure_Budget_plot} for $\eta=30\%$. 
On the 30th day, the actual budget is updated to $\$1078.88$, $\$1072.26$, and $\$1059.98$ using portfolio theory, FIC, and random weighting, respectively. 
As seen in this figure, the actual budget based on portfolio theory is higher than the actual budget based on fuzzy logic and random investment because it is based on optimization. The FIC, which considers both technical and fundamental features, results in invested budget very close to result of portfolio theory; suggesting that the FIC closely models the optimized behaviour of portfolio theory. 
In conclusion, as Fig. \ref{figure_Budget_plot} shows, both the proposed weight suggestion methods are better than random investment.


\section{Conclusion and Future Direction}

In this paper, a novel end-to-end artificial investment counselor was demonstrated which first forecasts the time series of prices in several stocks and then suggests the optimum portions of the budget to be invested in the stocks to have the best possible return in the future. 
The FIC, in contrast to portfolio theory, gives us the opportunity to consider even more features, such as news or sentimental attributes (e.g., social media impact \cite{yang2015twitter}), similar to an expert broker. 
Some future work can be considering more professional properties in finance, e.g., short selling and Elliott waves to have a more complete trading system.


\bibliographystyle{aaai}
\bibliography{references}

\end{document}